\documentclass{preprint}

\input tex.def
\input psfig.sty

\begin{document}

\title{What Can We Learn from Nearby AGNs?}
\author{Luis C. Ho}
\address{The Observatories of the Carnegie Institution of Washington \\
813 Santa Barbara Street \\
Pasadena, CA 91101-1292, USA}

\maketitle

\abstracts{
This contribution reviews the properties of nuclear activity in nearby 
galaxies, with emphasis on their implications for the demography of nuclear 
black holes and the nature of accretion flows in the regime of very low 
accretion rates.}

\section{Why Study Nearby AGNs?}

In a meeting largely devoted to surveys of distant, luminous AGNs, it 
is instructive to examine what we know about AGNs in very nearby galaxies.
Nearby AGNs are important for several reasons.  Since AGNs derive their 
power from accretion onto central black holes, the statistics of AGNs 
serve as a crude surrogate to delineate the demography of massive 
black holes in galaxies---a topic of considerable recent interest (see Ho 
2004a)---which is otherwise accessible only through painstaking kinematical 
observations.  With the growing realization that black holes, and thus 
AGN activity, are part and parcel of the life-cycle of many galaxies, there is
growing urgency to bridge the substantial gap that now exists between 
our knowledge of the space density of luminous quasars and that of 
quasar remnants.  The extant information on the luminosity function of AGNs 
fainter than, say, $M_B \approx -20$ to $-23$ mag, is very sketchy at 
virtually any redshift, being practically nonexistent for $M_B$ \gax\ $-18$ mag.
A census of nearby AGNs can place a robust constraint
on the very faint end of the local ($z\approx0$) AGN luminosity function.
By virtue of their proximity, nearby AGNs also offer a special vantage point to 
probe with high linear resolution the properties of their host galaxies and 
local environment.  This level of detail is indispensable, for example, if one 
wishes to understand the triggering or fueling mechanisms for AGNs.  Finally, 
as the evolutionary endpoint of quasars, nearby AGNs present an opportunity 
to study black hole accretion in a unique regime of parameter space, namely 
when the accretion drops to exceedingly low rates.  Much emphasis has been 
placed on how AGNs turn on; it is just as important to understand how 
they turn off.

\begin{figure}[t]
\psfig{file=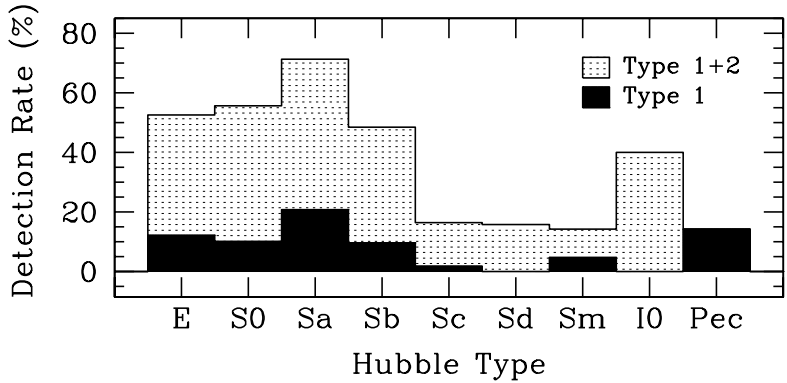,width=\textwidth,angle=0}
\caption{Detection rate of AGNs as a function of Hubble type in the
Palomar survey.  ``Type 1'' AGNs (those with broad H$\alpha$) are shown
separately from the total population (types 1 and 2).  (Adapted from Ho et al.
1997.)\label{fig1}}
\end{figure}

\section{Demography of Central Black Holes from AGN Surveys}

To the extent that an AGN signature signifies accretion onto a massive black
hole, a local AGN census gives us a lower limit on the fraction of nearby
galaxies hosting massive black holes (see Ho 2004b).  There are many ways to 
select AGNs.  Most surveys rely on selection criteria that isolate some 
previously known characteristics of AGNs.  A common strategy employs color 
cuts to highlight the blue continuum typically present in AGN spectra.  Some 
use objective prism plates to identify objects with strong and/or broad 
emission lines. Other wavelength-specific techniques to find candidate AGNs 
include preselection by radio or X-ray emission, or by ``warm'' infrared 
colors.  Variability is also occasionally used.  While all of these techniques 
have been successful, each introduces its own biases, and all ultimately 
require follow-up optical spectroscopy to confirm the AGN identification, to 
classify its type, and to determine its redshift.  

Nearby AGNs are generally intrinsically faint.  For these sources, most of 
the above-mentioned ``tell-tale signs'' are difficult or impossible to detect, 
either because they are severely diluted by the host galaxy or because they are
intrinsically absent.  To find local AGNs effectively, one has little recourse
but to resort to brute force: direct spectroscopy of a magnitude-limited 
sample of galaxies.  The following discussion draws mainly from the 
Palomar spectroscopic study of nearby galaxies (Ho, Filippenko, \& Sargent 
1997, and references therein), which remains the most sensitive 
survey of its kind.  The AGN samples emerging from the Sloan Digital Sky 
Survey (SDSS; Miller et al. 2003; Hao \& Strauss 2004; Kauffmann et al. 2004)
certainly eclipse the Palomar sample in size, and they provide much better 
statistics on sources with moderate and high luminosities, but they do not go 
as deep on the faint end of the luminosity function.  Moreover, as emphasized 
by Ho (2004b), the relatively large physical scale sampled by the SDSS
spectra complicates the interpretation of LINERs, locally the dominant 
constituent of the AGN population.

Figure 1 gives an overview of the AGN detection rate in the Palomar survey.
For conciseness, I do not distinguish the different subclasses of objects 
(Seyferts, LINERs, and transition objects), nor will I discuss the AGN 
``pedigree'' of each; these topics have been recently reviewed elsewhere (Ho 
2004b).   The most pertinent points to draw from the figure are:

\begin{itemize}
\item
AGNs are very common in nearby galaxies.  At least 40\% of all
galaxies (with $B_T \leq 12.5$ mag) emit AGN-like spectra.  

\item
The detectability of AGNs depends strongly on the morphological type of
the galaxy, being most common in early-type systems (E--Sbc).  The detection 
rate of AGNs reaches 50\%--75\% in ellipticals, lenticulars, and 
bulge-dominated spirals but drops to \lax 20\% in galaxies classified as Sc 
or later.

\item
These AGN detection rates support the notion, popularized by dynamical
studies, that central black holes are common in galaxies, and perhaps 
ubiquitous in those with bulges.  This consistency check is important, and 
the apparent agreement should not be taken for granted, because direct 
dynamical detections of black holes are technically challenging and still 
limited by small number statistics.
\end{itemize}


\section{Evidence for Intermediate-mass Black Holes}

The distribution of black hole masses based on direct dynamical 
measurements (see, e.g., Tremaine et al. 2002) currently does not extend below 
$\sim 3 \times 10^6$ \solmass, the mass of the central black hole in the 
Milky Way.  The record holder may be NGC 4945 ($\sim 1 \times 10^6$ \solmass; 
Greenhill, Moran, \& Herrnstein 1997), but the kinematics of its nuclear 
H$_2$O maser disk are not straightforward to interpret.  How far does the 
bottom end of the mass function of black holes extend?  Is there a sharp 
boundary between the masses of stellar and nuclear black holes?  The current 
limit of $\sim 10^6$ \solmass\ most likely reflects our present observational 
sensitivity rather than a physical threshold.  Intermediate-mass 
($\sim 10^3-10^5$ \solmass) black holes, if they exist, may offer important
clues to the nature of the ``seeds'' of supermassive ($10^6-10^9$ \solmass) 
black holes.  The inspiral of smaller black holes onto bigger ones, an 
inevitable consequence of hierarchical structure formation, may also provide a 
significant contribution to the integrated gravitational radiation background 
(e.g., Hughes 2002).

To date, the only known cases of intermediate-mass black holes based on 
resolved kinematics come from studies of globular clusters (Gebhardt, 
Rich, \& Ho 2002; Gerssen et al. 2002), which, in any case, have been 
controversial (Baumgardt et al. 2003a, b).  Clearly, it would be desirable to 
establish whether intermediate-mass black holes exist in galactic nuclei.  
Nuclear intermediate-mass black holes would manifest themselves as AGNs of 
moderately low luminosities (\lax $10^{43}$ \lum), most likely in low-mass 
or very late-type galaxies.

\begin{figure}[t]
\vbox{
\hbox{
\psfig{file=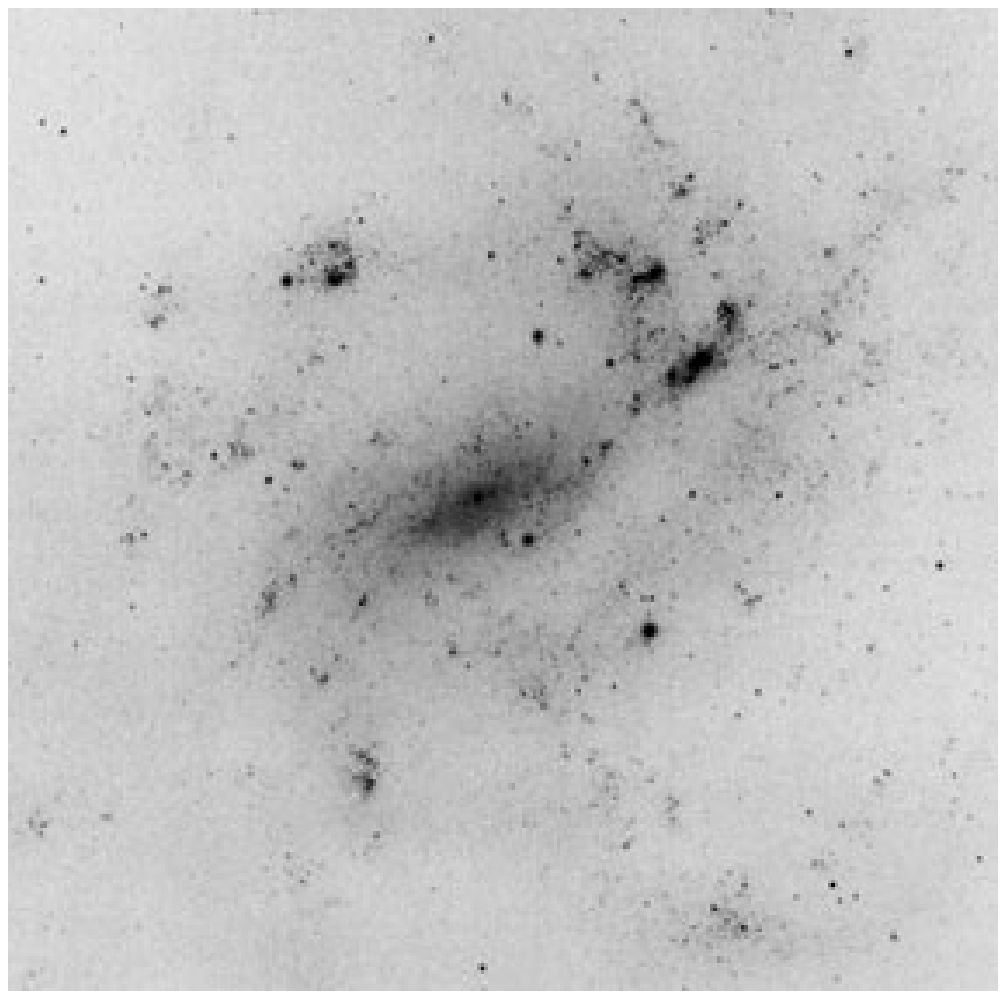,height=2.1truein,angle=0}
\psfig{file=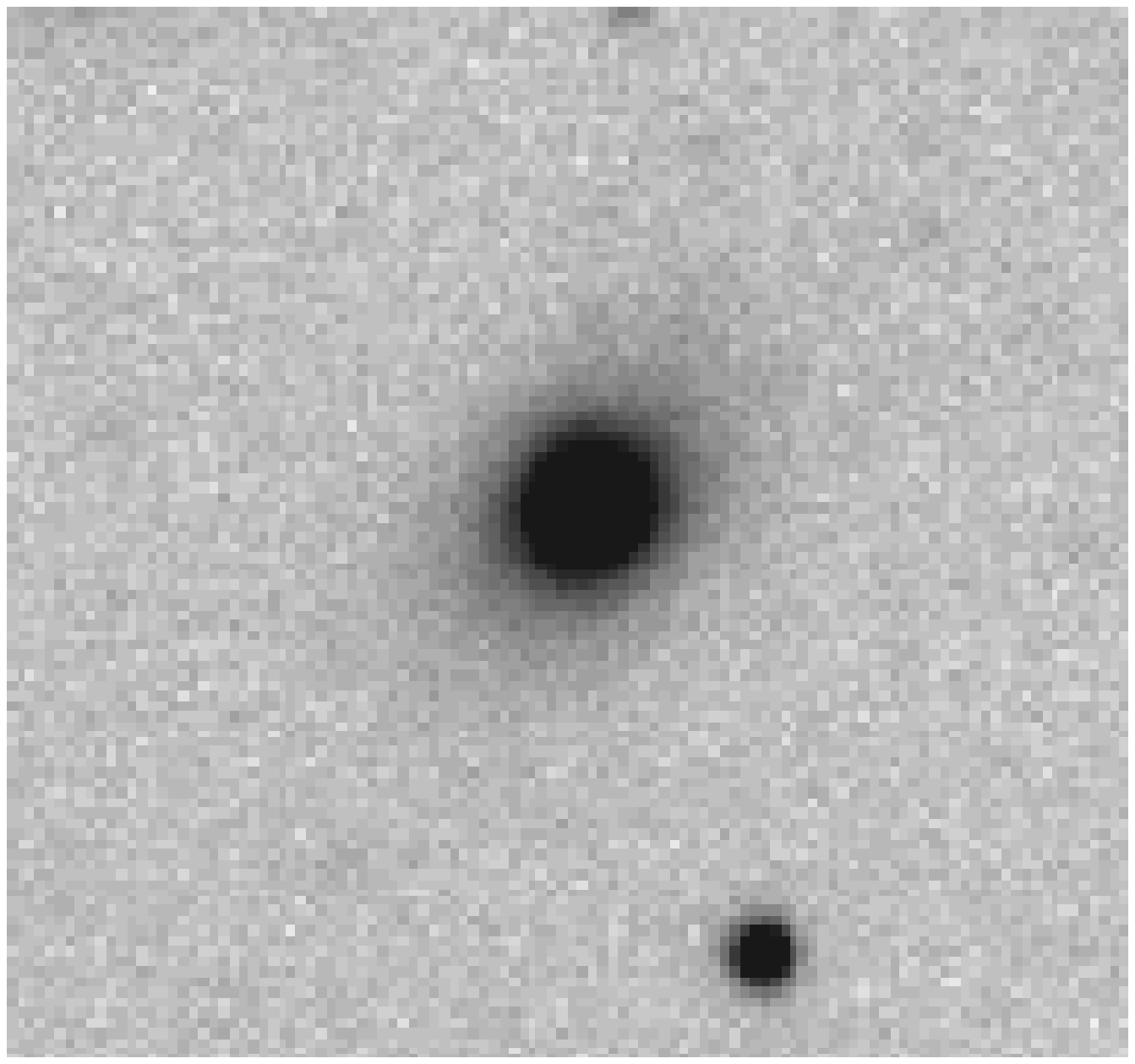,height=2.1truein,angle=0}
}}
\caption{
Two examples of AGNs in late-type galaxies.  The {\it left}\ panel shows an
optical image of NGC 4395, adapted from the Carnegie Atlas of Galaxies
(Sandage \& Bedke 1994); the image is $\sim$15\amin\ (17 kpc) on a side.  The 
{\it right}\ panel shows an $R$-band image of POX 52, adapted from Barth et 
al. (2004); the image is $\sim$25\asec\ (11 kpc) on a side.\label{fig2}}
\end{figure}

\begin{figure}[t]
\psfig{file=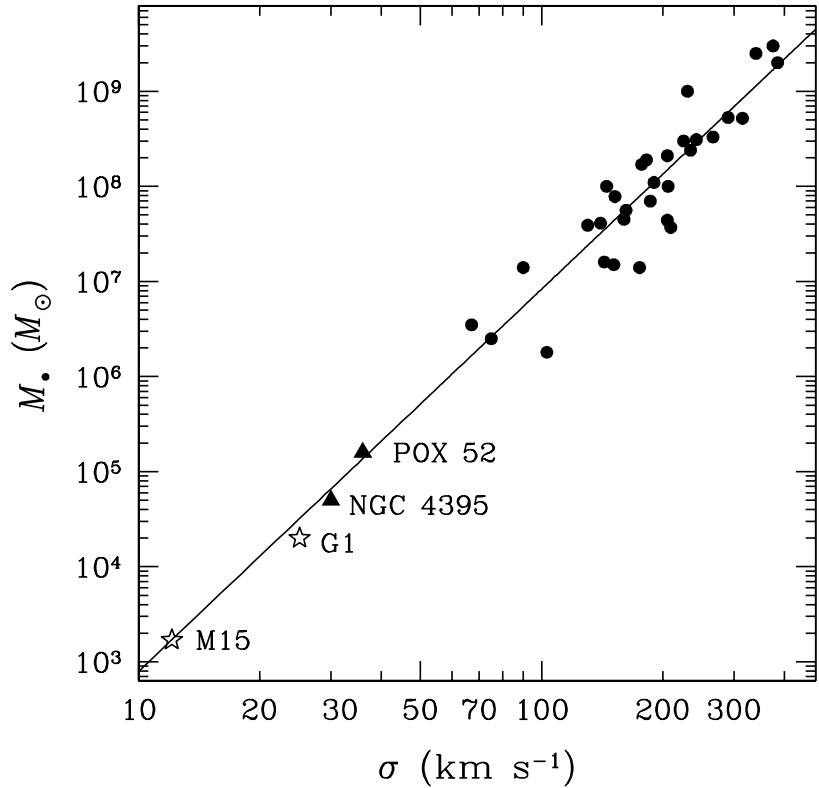,width=0.90\textwidth,angle=0}
\caption{
The black hole mass vs. velocity dispersion relation
extended to the regime of intermediate-mass black holes.  The dashed line
shows the fit of Tremaine et al. (2002): $M_\bullet \propto \sigma^{4.02}$.
\label{fig3}}
\end{figure} 

Two such cases have recently been reported (Fig. 2). The nearby ($\sim 4$ Mpc), 
Magellanic spiral (Sdm) galaxy NGC 4395 has long been known to host a 
low-luminosity Seyfert 1 nucleus (Filippenko \& Sargent 1989), which has been 
well studied from radio to X-ray wavelengths.  Filippenko \& Ho (2003) used 
several lines of evidence to argue that the mass of the central black hole in 
NGC 4395 lies in the range of $10^4 - 10^5$ \solmass.  This result is 
significant because it demonstrates unequivocally that nuclear black holes 
{\it can} exist in a {\it bulgeless}\ galaxy.  

Equally striking is POX 52, 
a considerably more distant ($\sim 90$ Mpc) dwarf galaxy.  Barth et al. (2004)
find that POX 52 has an optical spectrum that is virtually indistinguishable 
from that of NGC 4395.  It also shows tentative evidence of X-ray emission.
Based on the velocity width of the broad Balmer lines and the continuum 
luminosity, Barth et al. obtain a virial mass of $\sim 10^5$ \solmass.  
Interestingly, POX 52 appears to a dwarf elliptical galaxy, the first known 
to host an unambiguous AGN.  This is quite unexpected because dwarf elliptical 
galaxies, while technically spheroidal systems, bear little physical 
resemblance to classical bulges.  Dwarf ellipticals occupy a distinct locus on 
the fundamental plane of hot stellar systems (Bender, Burstein, \& Faber 1992; 
Geha, Guhathakurta, \& van~der~Marel 2003), and they are thought to originate
from harassment and tidal stripping of late-type (bulgeless) disk galaxies 
(e.g., Moore et al. 1996).  Thus, like NGC 4395, POX 52 stands as testimony 
that a bulge is not a prerequisite for the formation of central black hole.

Finally, another surprise (Fig. 3).  Although no firm conclusions can yet be 
drawn based on such meager statistics, it is intriguing, indeed mildly 
perplexing, that the four new candidate intermediate-mass black holes (the 
globular clusters M15 and G1 and the AGNs in NGC 4395 and POX 52) evidently 
seem to obey the relation between black hole mass and stellar velocity 
dispersion established by supermassive black holes (Gebhardt et al. 2000; 
Ferrarese \& Merritt 2000; Tremaine et al. 2002).

Are AGNs in dwarf or late-type galaxies common?  Evidently not, at least 
in the nearby Universe.  NGC 4395 is one of a kind in the Palomar survey of
$\sim$500 galaxies.  On the other hand, POX 52 was discovered in an objective 
prism 
survey of a relatively small area of sky (Kunth, Sargent, \& Bothun 1987), so 
objects like it cannot be that rare.  Using the first data release from the 
SDSS, Greene \& Ho (2004) have identified $\sim$150 AGNs that have sub-$10^6$ 
\solmass\ black holes.  The SDSS images do not furnish reliable 
information on the host galaxy morphologies, but a sizable fraction of 
them appear to be relatively disk-dominated spirals.

\section{The Optical Luminosity Function of $z \approx 0$ AGNs}

Many astrophysical applications of AGN demographics benefit from knowing
the AGN luminosity function, $\Phi(L,z)$.  Whereas $\Phi(L,z)$ has been 
reasonably well charted for high $L$ and high $z$ using quasars, it is very 
poorly known at low $L$ and low $z$.  Indeed, until very recently there has 
been no reliable determination of $\Phi(L,0)$.  Since nearby AGNs are faint, 
disentangling the nuclear emission from the much brighter contribution 
of the host galaxy poses a major challenge.  It is unacceptable to simply 
use the integrated emission from the entire galaxy.
Moreover, most optical luminosity 
functions of bright, more distant AGNs are specified in terms of the 
nonstellar optical continuum (usually in the $B$ band), whereas spectroscopic 
surveys of nearby galaxies generally only reliably measure optical line 
emission (e.g., H\al) because the featureless nuclear continuum is often 
impossible to detect in ground-based, seeing-limited apertures.  

\begin{figure}[t]
\psfig{file=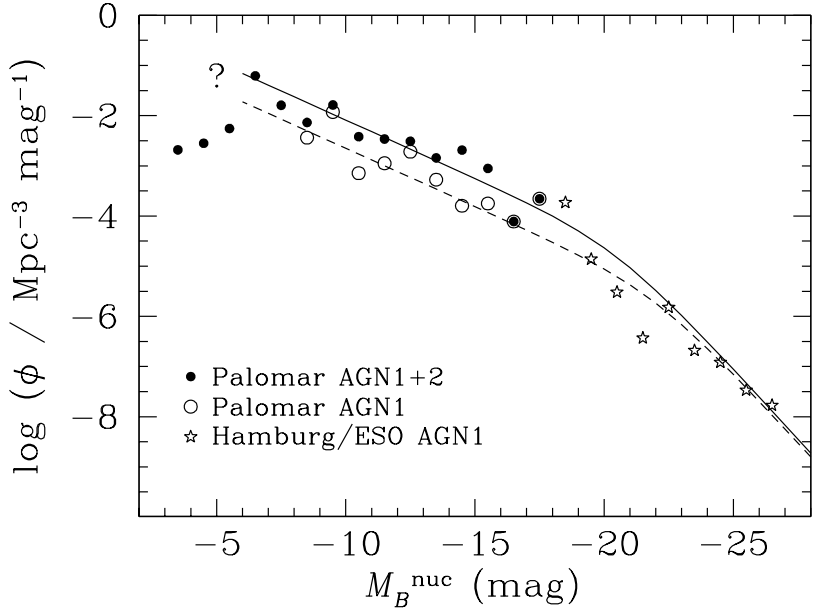,width=\textwidth,angle=270}
\caption{
The $B$-band nuclear luminosity function of nearby AGNs derived
from the Palomar survey.  The filled circles include all (type~1 + type~2)
sources, while the open circles include only type~1 sources.  The
sample of luminous Seyfert~1s and quasars from the Hamburg/ESO survey of  
K\"ohler et al. (1997) is shown as stars.  A double power-law fit to the 
Palomar and Hamburg/ESO samples is shown as a solid (type~1 + type~2) and
dashed (type~1) curve.  (Adapted from Ho 2004b.) \label{fig4}}
\end{figure}

The strategy adopted for the Palomar survey utilizes the well-known 
correlation between Balmer emission-line luminosity and optical featureless 
continuum luminosity, which has been shown by Ho \& Peng (2001) to hold 
for low-luminosity AGNs.  Figure 4 shows the $B$-band nuclear luminosity 
function for the Palomar AGNs.  Two versions are presented, each representing 
an extreme view of what kind of sources should be regarded as {\it bona fide}\ 
AGNs.  The open circles include only type~1 nuclei, sources in which broad 
H\al\ emission was detected and hence whose AGN status is incontrovertible.  
This may be regarded as the most conservative assumption and a lower bound, 
since genuine narrow-lined AGNs do exist. The solid circles 
lump together all sources classified as LINERs, transition objects, or 
Seyferts, both type~1 and type~2.  This represents the most optimistic view 
and an upper bound, since undoubtedly some narrow-lined sources must be 
stellar in origin but masquerading as AGNs.  The true space density of local 
AGNs most likely lies between these two possibilities.  In either case, the 
differential luminosity function is reasonably well approximated by a single 
power law from $M_B\,\approx$ --5 to --18 mag, roughly of the form 
$\Phi \propto L^{-1.2\pm0.2}$.  The slope may flatten for $M_B$ \gax\ --7 mag, 
but the luminosity function is highly uncertain at the faint end because of 
density fluctuations in our local volume.

For comparison, I have overlaid the luminosity function of $z$ \lax\ 0.3 
quasars and Seyfert 1 nuclei as determined by K\"{o}hler et al. (1997) from the
Hamburg/ESO UV-excess survey, scaled to the adopted cosmological 
parameters of $H_0$ = 75 \kms\ Mpc$^{-1}$, $\Omega_{\rm m} = 0.3$, and 
$\Omega_{\Lambda} = 0.7$.  This sample extends the luminosity function from 
$M_B \,\approx$ --18 to --26 mag.  Although the two samples do not strictly 
overlap in luminosity, it is apparent the two samples roughly merge, and that 
the break in the combined luminosity function most likely falls near 
$M_B^* \approx -19$ mag, where the space density 
$\phi \approx 1\times 10^{-4}$ Mpc$^{-3}$ mag$^{-1}$.  

\section{Radiative Inefficient Accretion}

As Figure 4 shows, clearly the luminosities of most nearby nuclei are 
extremely low.  To cast this statement in more physical terms, I have 
converted the optical luminosities to bolometric luminosities and 
then compared them relative to the Eddington luminosities, which were
estimated from the black hole mass vs. velocity dispersion relation (Fig. 5).
Note that nearly all the objects have $L_{\rm bol} < 10^{44}$ \lum, and most 
significantly less. Seyferts are on average 10 times more luminous than LINERs 
or transition objects.  More importantly, nearby nuclei are highly 
sub-Eddington systems.  All objects have $L_{\rm bol}/L_{\rm Edd} < 1$, with 
most $< 10^{-3}$.  One might argue that this is to be expected.  After all, 
in the present-day Universe bulge-dominated galaxies have spent most of 
their gas supply, leaving little fuel for central accretion.  But this is not 
the whole story.

It is true that not much fuel is needed to sustain the low level of activity 
observed.  For a canonical radiative efficiency of $\eta = 0.1$ 
appropriate for a geometrically thin disk, $L_{\rm bol} 
= 10^{40}-10^{42}$ \lum\ requires only $\dot M = 10^{-6}-10^{-4}$ \solmass\ 
yr$^{-1}$, ostensibly a minuscule amount.  The trouble is that the 
innermost regions of early-type galaxies should have much larger gas 
reservoirs than this.  Ho (2004, in preparation) estimates that most 
galactic nuclei should have gas supplies as large as $\dot M \approx 10^{-3}$ 
\solmass\ yr$^{-1}$, predominantly coming from mass loss from evolved stars 
and Bondi accretion of hot gas.  If this material were to be accreted and 
converted to radiation with $\eta = 0.1$, nearby AGNs should be much more 
spectacular that observed.  This suggests that either only a tiny fraction of 
the available gas gets accreted or that $\eta$ is much less than 0.1, as 
postulated in models of radiatively inefficient accretion flows (see Quataert 
2001 for a review).  If the gas is prevented from accreting, it is unlikely 
that supernova winds are the culprit, since there is little evidence of recent 
star formation in nearby nuclei (Ho et al. 2003).  Instead, recent models of 
radiatively inefficient accretion flows suggest that these systems are 
naturally prone to generating winds or outflows, which would curtail the 
amount of material that actually gets accreted.  Note, however, that this does 
not obviate the need for $\eta$ to be small---only that it does not 
have to be as small as it would be in the absence of winds---because 
a radiatively inefficient flow is a precondition for establishing the winds.

Nearby, low-luminosity AGNs are not simply scaled-down versions of luminous 
AGNs.  As a direct consequence of their low accretion rates, most nearby AGNs 
do not have ``standard'' optically thick, geometrically thin accretion disks.
Instead, their central engines are more akin to the class of radiatively 
inefficient accretion flows discussed in the recent literature.
Additional arguments in support of this picture can be found in Ho (2003).

\begin{figure}[t]
\psfig{file=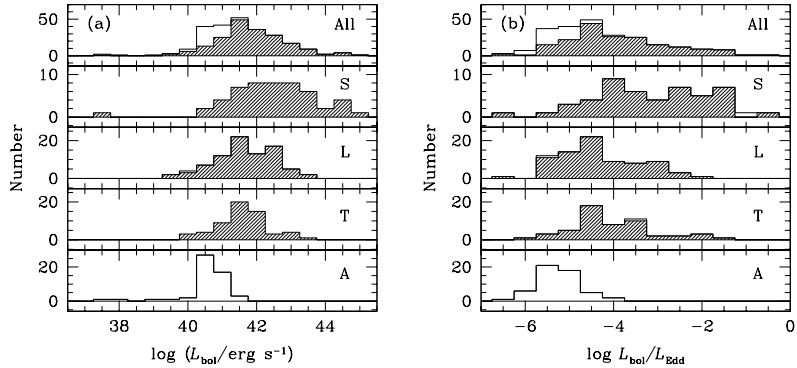,width=\textwidth,angle=270}
\caption{
Distribution of ({\it a}) nuclear bolometric luminosities
and ({\it b}) Eddington ratios $\lambda\,\equiv\,L_{\rm bol}/L_{\rm Edd}$.
S = Seyferts, L = LINERs, T = transition objects, and A = absorption-line
nuclei.  Open histograms denote upper limits.  (From Ho 2004b.) \label{fig5}}
\end{figure}

%


\end{document}